\documentclass[a4paper,fleqn,usenatbib]{mnras}

\usepackage{mathptmx}
\usepackage[T1]{fontenc}
\usepackage{ae,aecompl}

\usepackage{graphicx}	% Including figure files
\usepackage{amsmath}	% Advanced maths commands
\usepackage{amssymb}	% Extra maths symbols
\usepackage[caption=false]{subfig}
\usepackage{color}
\title[]{ALMA observations of cold molecular gas in AGN hosts at
  $z\sim$1.5 - Evidence of AGN feedback?}

\author[D. Kakkad et al.]{
D. Kakkad$^{1,2}$\thanks{E-mail: dkakkad@eso.org},
V. Mainieri$^{1}$,
M. Brusa$^{3,4}$,
P. Padovani$^{1}$,
S. Carniani$^{5,6}$,
C. Feruglio$^{7}$,
\newauthor ~M. Sargent$^{9}$,
B. Husemann$^{1,10}$,
A. Bongiorno$^{8}$,
M. Bonzini$^{1}$,
E. Piconcelli$^{8}$,
\newauthor ~J. D. Silverman$^{11}$,
W. Rujopakarn$^{11,12}$
\\
$^{1}$European Southern Observatory, Karl-Schwarzschild-Str. 2, 85748, Garching bei M\"unchen, Germany\\
$^{2}$Ludwig Maximilian Universit\"{a}t, Professor-Huber-Platz 2, 80539, M\"{u}nchen, Germany\\
$^{3}$Dipartimento di Fisica e Astronomia, Universit\`{a} di Bologna, viale Berti Pichat 6/2, I-40127 Bologna, Italy\\
$^{4}$INAF - Osservatorio Astronomico di Bologna, via Ranzani 1, I-40127 Bologna, Italy\\
$^{5}$Cavendish Laboratory, University of Cambridge, Madingley Road, Cambridge CB3 0HA, UK\\
$^{6}$Kavli Institute for Cosmology, University of Cambridge, Madingley Road, Cambridge CB3 0HA, UK\\
$^{7}$INAF Osservatorio Astronomico di Trieste, Via G. Tiepolo 11 , I-34124 Trieste\\
$^{8}$INAF - Osservatorio Astronomico di Roma, via Frascati 33, 00040 Monteporzio Catone, Italy\\
$^{9}$Astronomy Center, Department of Physics and Astronomy, University of Sussex, Brighton, BN1 9QH, UK\\
$^{10}$Max-Planck-Institut f\"ur Astronomie, K\"onigstuhl 17, D-69117 Heidelberg, Germany\\
$^{11}$ Kavli Institute for the Physics and Mathematics of the Universe, Todai Institutes for Advanced Study, \\ ~~the University of Tokyo, Kashiwa, Japan 277-8583 (Kavli IPMU, WPI)\\
$^{12}$Department of Physics, Faculty of Science, Chulalongkorn University, 254 Phayathai road, Pathumwan, Bangkok 10330, Thailand.
}

\date{Accepted ???. Received ???; in original form ???}

\pubyear{2016}

\begin{document}
\label{firstpage}
\pagerange{\pageref{firstpage}--\pageref{lastpage}}
\maketitle
\begin{abstract}
Similarly to the cosmic star formation history, the black hole accretion rate density of the Universe peaked at 1$<z<$3. This cosmic epoch is hence best suited for investigating the
    effects of radiative feedback from AGN. Observational efforts are underway to quantify the impact of AGN feedback, if any, on their host galaxies. Here we present a study of the molecular
    gas content of AGN hosts at z$\sim$1.5 using CO[2-1] line emission
    observed with ALMA for a sample of 10 AGNs. We compare this with a sample of 
    galaxies without an AGN matched in redshift, stellar mass, and star formation rate. We detect CO in 3 AGNs with
    {$\mathrm{L_{CO} \sim 6.3-25.1\times 10^{9} L_{\odot}}$ which
      translates to a molecular hydrogen gas mass of
      $\mathrm{2.5-10\times 10^{10} M_{\odot}}$} assuming conventional
    conversion factor of $\mathrm{\alpha_{CO}}\sim3.6$. Our results indicate a >99\% probability of lower depletion time scales and lower molecular gas fractions in AGN hosts with respect to the non-AGN comparison sample. We discuss the implications of these observations on the impact that AGN feedback may have on star formation efficiency of z>1 galaxies.
\end{abstract}


%%%%%%%%%%%%%%%%%%%%%%%%%%%%%%%%%%%%%%%%%%%%%%%%%%

%%%%%%%%%%%%%%%%% BODY OF PAPER %%%%%%%%%%%%%%%%%%
\section{Introduction} \label{sect:introduction}

In the past decade, using mm and sub-mm telescopes there has been a 
remarkable progress in the study of cold molecular gas content in galaxies.
\citep[e.g.][]{greve05,daddi10,engel10,
  tacconi10,davis11,geach11,bauermeister13, bothwell13,saintonge13,
  tacconi13, sargent14, genzel15}. This information provides a key
ingredient to galaxy evolutionary studies as it is out of this gas
that the galaxy may form stars.

Most of the molecular gas in the interstellar medium (ISM) exists in
the form of molecular hydrogen ($\mathrm{H_{2}}$). However, in the
commonly used optical or radio portions of the electromagnetic
spectrum, $\mathrm{H_{2}}$ does not have any emission or absorption
lines at the typical cool temperatures of the ISM, which makes it
necessary to use other molecules as its tracers in most instances. The
high abundance of Carbon Monoxide (CO) makes it one of the most
popular tracer, although other molecules have also
often been used in the literature \citep[e.g.][]{gao04,
  greve05,alaghband13, gullberg15, heyer15}.  The collisions of the
tracer molecules with $\mathrm{H_{2}}$ results in emission of photons
due to excitation or de-excitation. The emission from these tracer molecules fall in wavelength regions which are more easily observed than those of $\mathrm{H_{2}}$. The conversion
between CO luminosity and $\mathrm{H_{2}}$ gas mass is usually
parametrized through a factor, $\mathrm{\alpha_{CO}}$, which typically
ranges between 0.8 and $\sim$4 (in units of 
${M_{\odot}}~[\rm K~km~s^{-1}~pc^2]^{-1}$) for galaxies with an ISM enriched to more or less the Solar abundance \citep[e.g.][]{arimoto96}.

Star forming galaxies form a tight correlation between their star
formation rate (SFR) and stellar mass ($\mathrm{M_{\ast}}$) in the
range of 9.5<log M$_{\ast}$/$\mathrm{M_{\odot}}$<11.5, the so-called
``main-sequence'' (MS) of star forming galaxies
\citep[e.g.][]{daddi07, elbaz07, noeske07, pannella09, whitaker12,
  kashino13, rodighiero14, speagle14}. The Main Sequence relation is known to
exist in a range of redshifts with an evolution in its normalization
\citep[e.g][]{whitaker12,bonzini15} as well as its slope
\citep[e.g.][]{speagle14,schreiber15}. The molecular gas properties of
normal galaxies on the MS and starburst (SB) galaxies lying above the MS
correlation have been the subject of several studies aiming at
quantifying the relation between the molecular gas content or gas
fractions of these galaxies with other galaxy properties such as the
SFR. Most of these studies are based on molecular gas measurements as
traced by CO. Results from previous work suggest that there are
in fact two modes of star formation in high redshift galaxies: a
starburst one with the molecular gas usually distributed in a compact
configuration and short consumption times ($\mathrm{10^{7}-10^{8}}$
years), typical of submillimetre galaxies (and local ultra-luminous
infrared galaxies [ULIRGs]); and a quiescent one, with gas reservoirs
distributed in extended disks and longer consumption times
($\mathrm{10^{9}}$ years), which is observed on the MS at $z>1$ (and
in local spiral galaxies) \citep{daddi10,genzel10,sargent14,silverman15}.

AGNs could play an important role in regulating the ISM content of
their host galaxies through radiative winds and mechanical energy,
which, if powerful enough, could have the capability to sweep the
entire galaxy clean of gas, a process generically named radiative mode of
``AGN feedback'' \citep[see review by][]{fabian12}. AGN feedback has
also been invoked to reduce the number of
massive galaxies, which otherwise is over-predicted by simulations
assuming $\Lambda$CDM cosmology \citep{silk12}. The redshift range 
1$<z<$3 presents an interesting laboratory to test radiative feedback
from AGNs as this epoch witnessed the peak accretion activity of black
holes and high star formation rate in the universe
\citep[e.g.][]{lilly96,shankar09, madau14}. There has been growing
evidence of powerful and extended kiloparsec scale outflows from these
high redshift quasars
\citep{harrison12,brusa15a,perna15,kakkad16,wylezalek16,
  zakamska16} which may be affecting their host galaxies by
suppressing star formation in regions dominated by the outflows
\citep{cano-diaz12, cresci15,carniani16}.

A promising way to quantify the impact that these AGN driven outflows 
may have on their host galaxies is by studying their cold molecular 
gas properties. Nowadays, a systematic study of the molecular gas 
properties of AGN host galaxies is still missing. 
Most of the molecular gas
studies in AGNs have been restricted to high luminosity AGNs which
usually lie above the main-sequence, in the so called starburst regime
whose molecular gas properties mimic those of the non-AGN starburst
population and local ULIRGs \citep[e.g][]{aravena08, polletta11,
  riechers11, villar-martin13, feruglio14, stefan15,
  daisy-leung16}. Some other studies have focused on
unobscured AGNs for which measurements of gas fractions are highly
uncertain owing to the inaccurate stellar mass determinations.

Also, there is very little knowledge about the molecular gas content
of obscured AGNs as we move to lower bolometric luminosities at
$z>1$, which coincides with the peak of the star formation and black
hole accretion activity in the universe.  \citet{mullaney15} showed
that most of these low-to-moderate luminosity AGNs lie on the MS of
the star forming galaxies which make them an ideal choice for
comparative study between AGNs and normal star forming
galaxies without an AGN. Throughout the paper, 
we define ``non-AGN'' galaxies as
those which lie on the MS of star forming galaxies and have an X-ray
luminosity (2-10 keV) less than $\mathrm{10^{42}}$ erg/s, 
while any object above this X ray luminosity is regarded as an AGN.

In this paper we present an investigation
of the molecular gas properties of 10 obscured AGNs at z$\sim$1.5 with the Atacama
Large Millimeter/sub-mm Array (ALMA) selected based on their position
on the MS of star forming galaxies. Our ultimate aim is to compare the
gas properties of the AGN sample with those of the non-AGN
star forming population to infer if there is a difference in the gas
content and consequently the star formation process between the two
groups.

This paper is arranged as follows: Section \ref{sect:target_selection}
mentions our strategy for selecting the AGN sample. Observation
parameters and details about the data reduction are summarized in
Sect. \ref{sect:observations_reductions}. Sect. \ref{sect:results}
describes the results obtained and their implications for the big
picture. Sect. \ref{sect:discussion} discusses our results, while 
in Sect. \ref{sect:summary} we present our
conclusions. Throughout this paper, we use the following cosmology:
$H_{0}$ = 70 km/s, $\Omega_{\Lambda}$ = 0.7, $\Omega_{m}$ = 0.3,
$\Omega_{r}$ = 0.0.

\section{Target selection} \label{sect:target_selection}

We focus our study on a sample of AGNs at $z\sim$1.5, close to the
epoch of peak accretion activity of black holes, ideal for testing
radiative AGN feedback. We chose AGNs with host galaxies on the MS
where most of the moderate luminosity AGN and non-AGN
host galaxies are located \citep{mullaney15}. The 
non-AGN galaxy sample
matched in M$_{\ast}$, SFR and $z$ forms a key component of this study
since it is with respect to this population that the molecular gas
properties of our AGN sample will be compared to. We restrict our
AGN sample to moderate luminosity
($\mathrm{L_{bol}\approx10^{43-46} erg/s}$) obscured AGNs which sets this
study apart from previous work, which has focused on high luminosity
AGNs ($\mathrm{L_{bol}>10^{46}erg/s}$) which usually lie above the
MS. Our selection criteria aim at probing the molecular gas properties 
of the general moderate luminosity AGN population.

Numerous parametrizations have been used for the MS of star forming
galaxies in the literature \citep[e.g.][]{pannella09, whitaker12,
  schreiber15}. Each MS locus has been derived from a different sample
and they suffer from many selection effects \citep{renzini15}. For the
purpose of this work, we adopt one of the most recent parametrization
from \citet{schreiber15} which uses a large sample of galaxies to
investigate the redshift evolution of the MS in the CANDELS
field. Their sample is mass complete above
$\mathrm{2\times 10^{10}~M_{\odot}}$ and is able to trace the higher
end of the MS slope more accurately compared to previous works.

\begin{figure*}
\centering
\includegraphics[width=5.0in,height=3.5in]{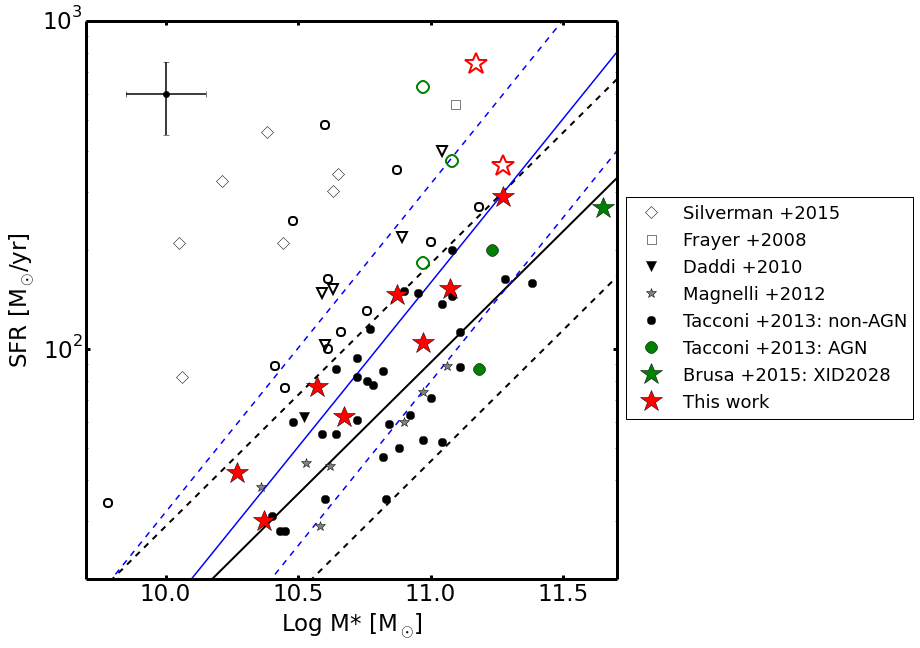}
\caption{Sample selection: the solid black line corresponds to the MS line for star forming galaxies adapted from \citet{schreiber15} with a scatter of 0.3 dex shown by the dotted black line. Galaxies falling on the MS are shown by filled symbols while those which fall above the MS are shown by open symbols. Coloured symbols represent the AGN sample while the black and gray symbols represent the non-AGN samples. The sample from the literature are from \citet{silverman15, frayer08, daddi10, magnelli12, tacconi13} and \citet{brusa15} (see legend). Our sample is shown by the red stars. The blue lines show the MS line  with scatter derived from \citet{pannella09}, the original basis of our sample selection.  The error bars at the top left corner of the plot show the representative errors for the data points ($\sim$0.3 dex on both axes.)
  \label{fig:sample_selection}}
\end{figure*}

We selected our AGN sample in the X-ray band which is almost free from
contamination by other sources such as normal star forming
regions. We started with $\sim$4500 AGNs merging the Chandra and
XMM-Newton surveys of the COSMOS field \citep{brusa10, civano12}, the
Chandra survey of the Extended Chandra Deep Field South
\citep[E-CDFS:][]{lehmer05}, and the 4Ms Chandra observations of the
Chandra Deep field South \citep[CDFS:][]{xue11}. All these survey
areas have extensive multi-wavelength coverage from the radio to the UV band, 
which allow a proper physical characterization of the galaxy properties
(e.g. stellar mass, star formation rate, morphology), \citep{tozzi01,
  szokoly04, mainieri05, brusa10, salvato11, civano12, bongiorno12,
  rosario12}. Out of this large X-ray sample, we considered only those
targets which have  high quality flag for secure spectroscopic redshift determination (Class 3 and 4 according to convention by \citet{lilly07}) and accurate SFR and
$\mathrm{M_{\ast}}$ estimates. The SFR were estimated from Herschel
Photoconductor Array Camera and Spectrometer (PACS) photometry using
the relation $\mathrm{SFR = 10^{-10}\cdot L_{IR} [L_{\odot}]}$
assuming a Chabrier initial mass function (IMF). With the Herschel
PACS detection, one can effectively impose a lower limit on the SFR
and consequently M$_{\ast}$. Based on the SFR and the AGN bolometric
luminosity, $\mathrm{L_{bol}<10^{46} erg s^{-1}}$, we do not expect
any significant contamination from the AGN to the Far-Infrared (FIR)
end of the spectrum \citep{mullaney12, rosario12}. The stellar masses
of the host galaxies were computed using a two component (galaxy+AGN)
SED fitting technique \citep{bongiorno12, bonzini13}, assuming a
Chabrier IMF.  Typical errors associated with the parameters such as SFR, $\mathrm{L_{bol}}$ and $\mathrm{M_{\ast}}$ due to the choice of SED templates used are about a factor of 2. We refer the reader to \citet{bonzini15} for details about the SED fitting. Our original selection based on the MS line derived by 
\citet{pannella09} returned a sample of 10 objects, of which C92 was an outlier in the
strong starburst regime (more in Sect. \ref{sect:results}). However,
with the new parametrization of the MS from \citet{schreiber15}, we
arrive at a final sample of 8 objects on the MS and 2 starburst
galaxies, shown by the red stars in
Fig. \ref{fig:sample_selection}. The basic properties of all the ALMA
targets are listed in Table \ref{table:target_properties}.

\begin{figure*}
\subfloat{\includegraphics[width=2.4in]{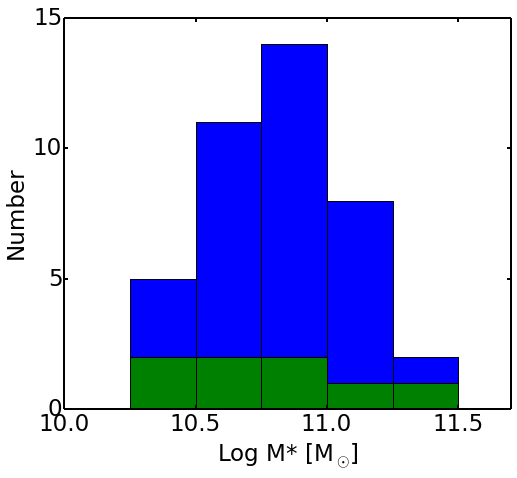}} 
\subfloat{\includegraphics[width=2.5in]{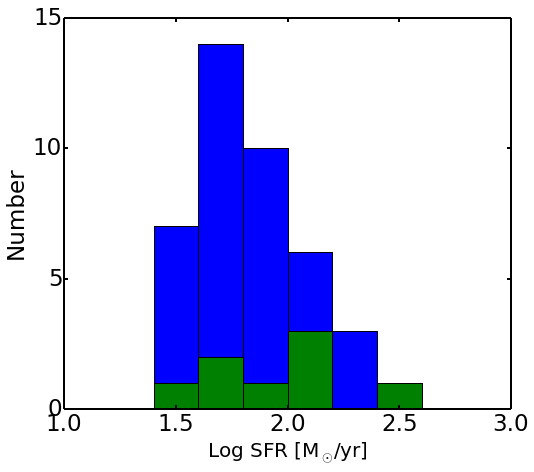}}
\caption{Histograms showing the $\mathrm{M_{\ast}}$ coverage (left) and SFR coverage (right) of the AGN sample presented in this paper (green) and the non-AGN sample (blue).
\label{fig:histograms}}
\end{figure*}

As already mentioned, a key element to this study is to have a
comparison sample of non-AGN galaxies on the MS. Since there is an
evolution in the properties of the galaxies and black holes with
redshift \citep[e.g.][]{walter16}, it is important to select our sample
matched in redshift. We constructed the comparison non-AGN sample
from the work of \citet{daddi10,tacconi13} and \citet{magnelli13} for
which the molecular gas measurements are available from CO
observations using IRAM-PdBI. We refer the reader to the corresponding
papers for further details about the observations and parent
sample. We selected only those objects falling in the redshift
range $1<z<2$ and on the MS at $z\sim$1.5. We also matched the
sample from these works in SFR and M$_{\ast}$ to that of the AGN
sample to make sure we are comparing similar galaxies. This is apparent from the histograms in Fig. \ref{fig:histograms}, which show the comparison between the $\mathrm{M_{\ast}}$ and SFR coverage of the AGN and non-AGN sample. All the SFR have been computed assuming a Chabrier IMF. Our
final AGN sample and all the other comparison samples in the
SFR-M$_{\ast}$ plane are shown in Fig. \ref{fig:sample_selection}. The
red stars correspond to the ALMA sample presented in this paper while
the black and grey filled symbols are the non-AGN galaxies. The open
symbols are the starburst galaxies, matched in M$_{\ast}$ and $z$, to
be used for analysis later to check if the properties of the AGN
sample are closer to those of the non-AGN star forming galaxies or
starburst galaxies. The MS line used for reference in this paper is
adapted from \citet{schreiber15} with a dispersion of 0.3 dex, details
of which were mentioned earlier in this section. The blue line
corresponds to the \citet{pannella09} MS parametrization, based on
which the sample was initially selected. All targets but C92 fall
under this parametrization (within 0.3 dex). It is apparent from the
figure that the AGN and the non-AGN galaxy samples cover the same
range in the SFR-M$_{\ast}$ plane. The same coverage in
$\mathrm{M_{\ast}}$ is highly desirable given that observations at
both low and high redshift galaxies point to an increasing gas
fraction with decreasing stellar mass \citep[e.g.][]{magdis12,
  saintonge12,tacconi13}.

We have complemented our ALMA main sequence AGN sample with literature AGN data
(green filled symbols in Fig. \ref{fig:sample_selection}). Five
galaxies from the \citet{tacconi13} sample have been classified as AGN
by matching them with the \citet{nandra15} AEGIS-X catalog  with X ray luminosity, $\mathrm{L_{2-10 keV} \sim 10^{42-43.5} erg/s}$. Two of
them are MS galaxies. XID2028 from \citet{brusa15} is also a main sequence
AGN. This source has been extensively studied in ionized
\citep{perna15,cresci15} as well as molecular \citep{brusa15a} gas
phases and has been shown to be in an active outflowing phase.

\begin{table*}
\centering
\begin{tabular}{ccccccccc}
\hline
ID & RA & DEC & $z^{a}$ & Log $\mathrm{L_{bol}}^{b}$ & Log M$_{\ast}^{c}$ & Log $\mathrm{L_{IR}}^{d}$ & SFR$^{e}$ & FIELD\\
 & (h:m:s) & ($^{\circ}:^{\prime}:^{\prime\prime}$) & & (erg/s) & ($\mathrm{M_{\odot}}$) & (L$_{\odot}$) & (M$_{\odot}$/yr) & \\
\hline
DETECTIONS & & & & & & & & \\
C1591* & 10 01 43 & +02 33 31 & 1.238 & 45.7 & 11.3 & 12.6 & 362 & COSMOS\\
C92*   & 10 01 09 & +02 22 55 & 1.581 & 45.4 & 11.2 & 12.8 & 740 & COSMOS\\
\#226   & 03 32 16 & $-$27 49 00 & 1.413 & 42.8 & 10.9 & 12.2 & 146 & CDFS\\
NON-DETECTIONS & & & & & & & & \\
X5308 & 09 59 22 & +01 36 18 & 1.285 & 45.5 & 11.0 & 12.0 & 104 & COSMOS\\
X2522 & 09 57 28 & +02 25 42 & 1.532 & 46.3 & 11.3 & 12.5 & 290 & COSMOS\\
C1148 & 10 00 04 & +02 13 07 & 1.563 & 46.8 & 11.1 & 12.2 & 153 & COSMOS\\
C488  & 10 01 47 & +02 02 37 & 1.171 & 45.4 & 10.3 & 11.6 & ~42 & COSMOS\\
C152  & 10 00 39 & +02 37 19 & 1.188 & 45.5 & 10.4 & 11.5 & ~30 & COSMOS\\
C103  & 10 01 10 & +02 27 17 & 1.433 & 45.1 & 10.7 & 11.8 & ~62 & COSMOS\\
\#682   & 03 32 59 & $-$27 45 22 & 1.155 & 43.4 & 10.6 & 11.9 & ~77 & CDFS\\
\hline
\end{tabular}
\caption{Target properties: \newline
$^{a}$Secure spectroscopic redshifts obtained from optical spectra,
\newline$^{b}$AGN bolometric luminosities obtained as presented in \citet{lusso11}, 
\newline$^{c}$Host galaxy stellar mass obtained with a two component (AGN+galaxy) SED fit assuming a Chabrier IMF \citep{bongiorno12, bonzini13}, 
\newline$^{d}$Total infra-red luminosity (8-1000$\mu$m) derived using Herschel PACS+SPIRE photometry, \newline$^{e}$Host galaxy star formation rate computed fitting the UV-to-FIR (Herschel) photometry with the \citet{berta13} template library and converting the best fit 8-1000$\mu$m luminosity to a SFR using the \citet{kennicutt98} prescription assuming a Chabrier IMF. 
\newline NOTE: Typical uncertainty in quantities such as $\mathrm{L_{bol}, M_{\ast}}$ and SFR, taking into account different SED templates used, is about 0.3 dex.
\newline*These galaxies fall under the SB regime in SFR-M$_{\ast}$ plane according to the \citet{schreiber15} MS parametrization.
  \label{table:target_properties}}
\end{table*}

\section{Observations and Data analysis} \label{sect:observations_reductions}

The ALMA observations of the 10 AGN host galaxies were carried out in
cycle 2 (Project code: 2013.1.00171.S, PI: V. Mainieri) using 35-38
12m antennas. The velocity/spectral resolution for all observations
was $\sim$24 km/s. The observations were carried out in Band 3, with the
receiver tuned to the frequency of
$\mathrm{^{12}CO(2-1)}$ emission (from now simply CO(2-1) unless
otherwise specified, rest frame frequency = 230.5 GHz). The CO(2-1)
transition was targeted in order to sample the extended, cool gas
reservoir of our galaxies which is assumed to contain the bulk of the
gas mass, thereby avoiding the uncertain excitation corrections
affecting higher-J transitions. The observations were executed in
three ``scheduling blocks'' to optimize the use of ALMA time by
simultaneously detecting CO(2-1) for different objects with the same
correlator configuration. The on-source exposure time was kept between
$\sim$2.5-8 minutes depending on the sensitivity requirements. The root-mean-square
(rms) noise requirements were calculated in order to obtain a 3$\sigma$
detection of the CO line assuming the $\mathrm{L_{CO}-L_{IR}}$
correlation as measured for MS galaxies in \citet{daddi10}. We reached
baselines up to 800 m resulting in beam sizes of
$\sim 1.5^{\prime\prime} \times 1.4^{\prime\prime}$ and
$0.8^{\prime\prime} \times 0.5^{\prime\prime}$. All the observed
properties of each target are reported in Table
\ref{table:observed_target_properties}. Summarizing, we have 3 CO
detections and 7 non-detections.

We used the calibrated data products as delivered to us, while imaging
for all the targets was done using CASA (version 4.2.2). The images
were cleaned using the CASA task ``CLEAN'' with ``natural'' WEIGHTING. The spectral binning was kept below 100
km/s in order to get at least 4 channels for CO detection, assuming an
average width of the CO(2-1) line of 400 km/s \citep{daddi10}. The
exact choice of the width depended on where we get the maximum S/N
ratio. Following this approach, imaging was done with velocity bins
of 50 km/s for C92 and for the remaining targets we used 80 km/s. 
The imaging results are described in
Sect. \ref{sect:results}. The angular scale of the final image is
160$^{\prime\prime}$ and the analysis was restricted within
25$^{\prime\prime}$ of the field center.

To determine the rms values, we used the CASA task "immoments" to
construct moment 0 maps by collapsing along the channels with CO
detection (for C92, C1591 \& \#226) or along the channels with
expected CO detection and computed rms on these maps. We did not
compute rms directly from the spectrum since the noise is a function
of frequency and this introduces systematic
uncertainties in our measurements \citep{maiolino15}. For galaxies not
  detected in CO, the CO flux $\mathrm{I_{CO}}$ in Table
  \ref{table:observed_target_properties} represents a 3-$\sigma$ upper
  limit calculated from sensitivity reached over a channel width of
  320 km/s around the expected position of CO(2-1) line. The choice
of 320 km/s velocity bin corresponds to the mean of the FWHM of CO(2-1)
line from the CO detections (see also 
Table \ref{table:fit_results}). Most of the undetected galaxies lie on the lower mass end of the AGN host mass distribution. Consequently, dynamical arguments suggests that we would expect a lower line width compared to the detected ones which are on the massive end of this distribution. This would lead to an expectation of lower CO luminosity than calculated using a width of 320 km/s based on previous reports on CO luminosity and linewidth relation in literature \citep[e.g.][]{bothwell13, carilli13, sharon16}. For 
C1591, C92 and \#226, we reached a sensitivity of 0.21, 0.13 and 0.11 
Jy/beam in 880 km/s, 320 km/s and 320 km/s wide spectral channels 
(these bins span the entire CO(2-1) line).

In case of a detection, the spectra were extracted from the ALMA cube
using a circular aperture with a radius of $\approx$1.5"
  in case of C1591 and C92 and $\approx$1" in case of \#226. We fit the CO(2-1) line using the IDL routine
MPFIT \citep{markwardt12} using a single Gaussian component, through
which the flux of the CO line and the errors on the fits were estimated. The continuum remained
undetected for all targets. The CO flux was also measured by fitting
the data in UV space with the CASA task ``uvmodelfit'' using a point
source or an elliptical Gaussian model which gives the centroid of the
line emission and the integrated flux. The line flux obtained with the
fitting routine and the UV model fit were compared with each other and
they agree well at one-sigma error levels.

\begin{figure*}
\subfloat{\includegraphics[width=2.2in]{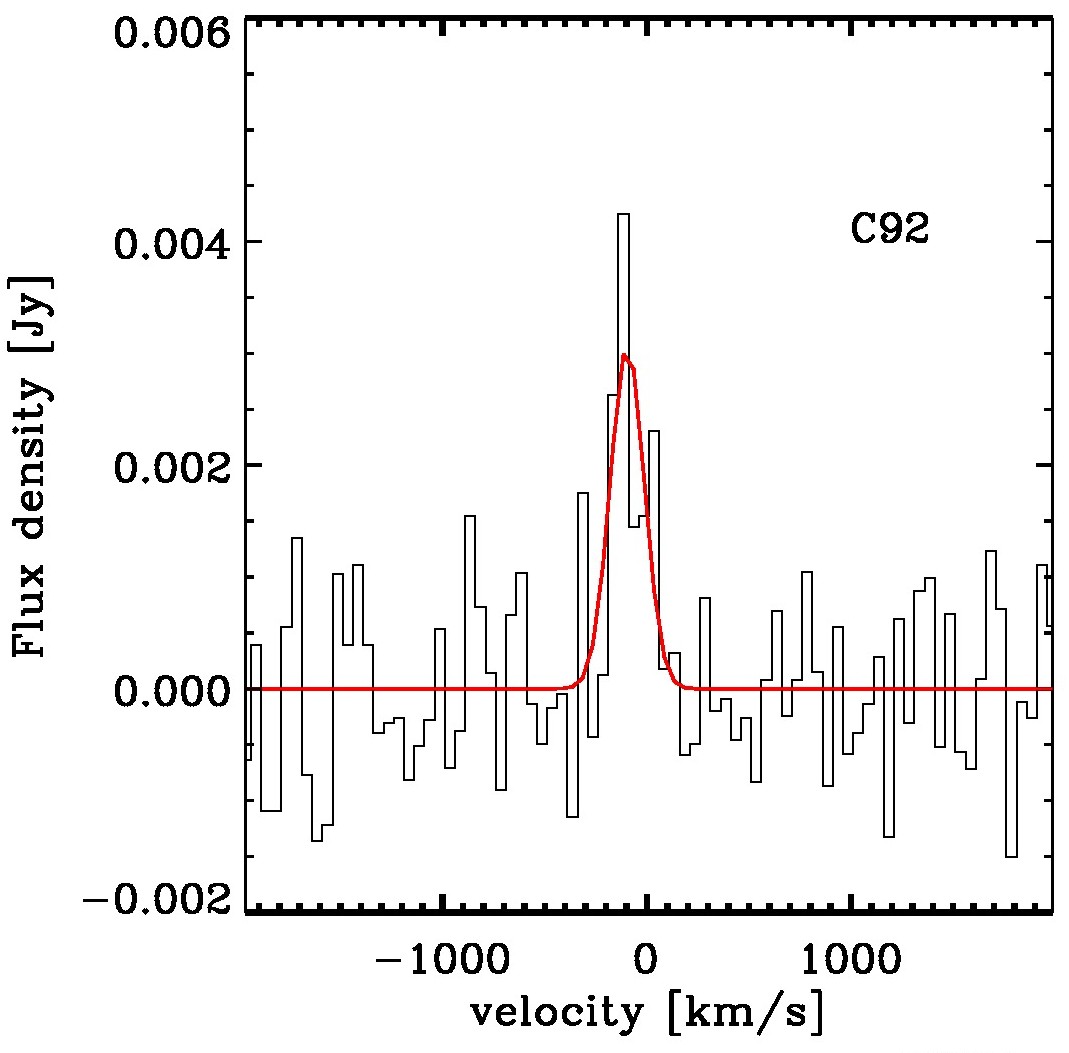}} 
\subfloat{\includegraphics[width=2.06in]{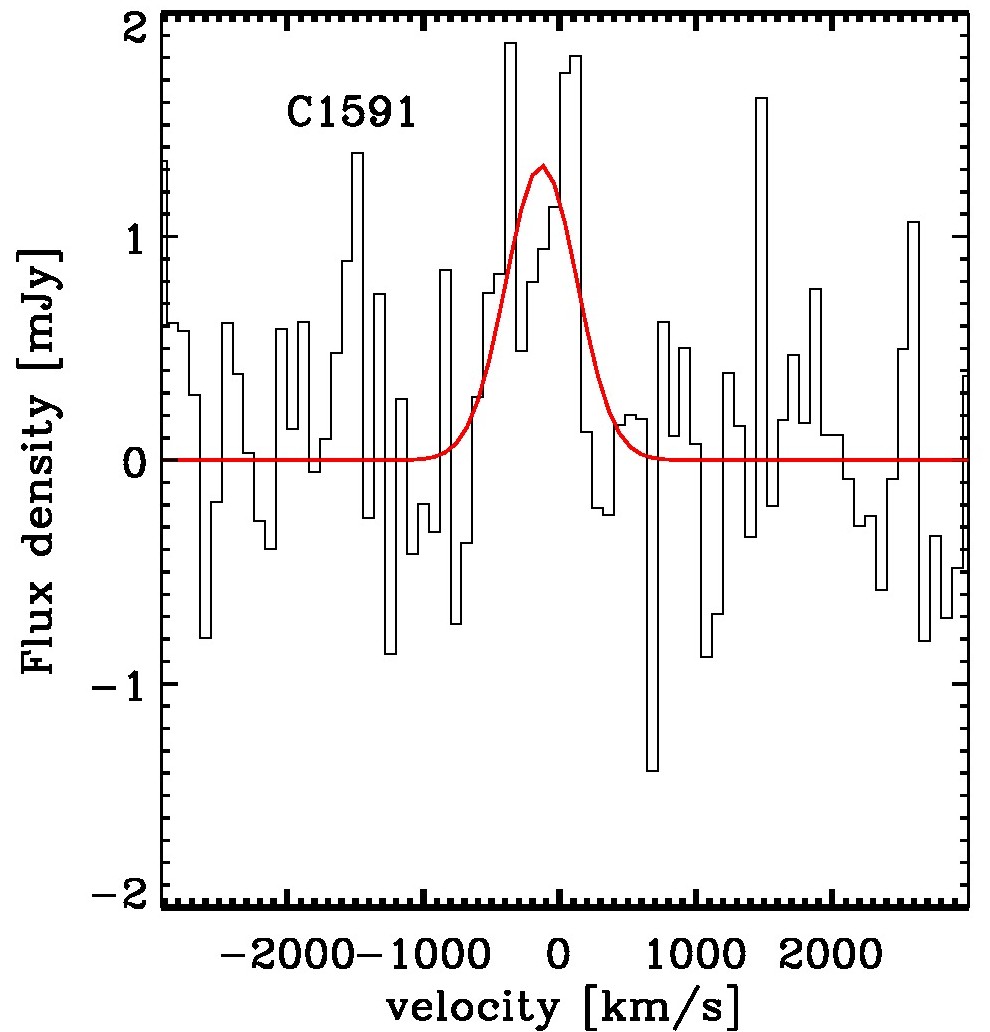}}
\subfloat{\includegraphics[width=2.17in]{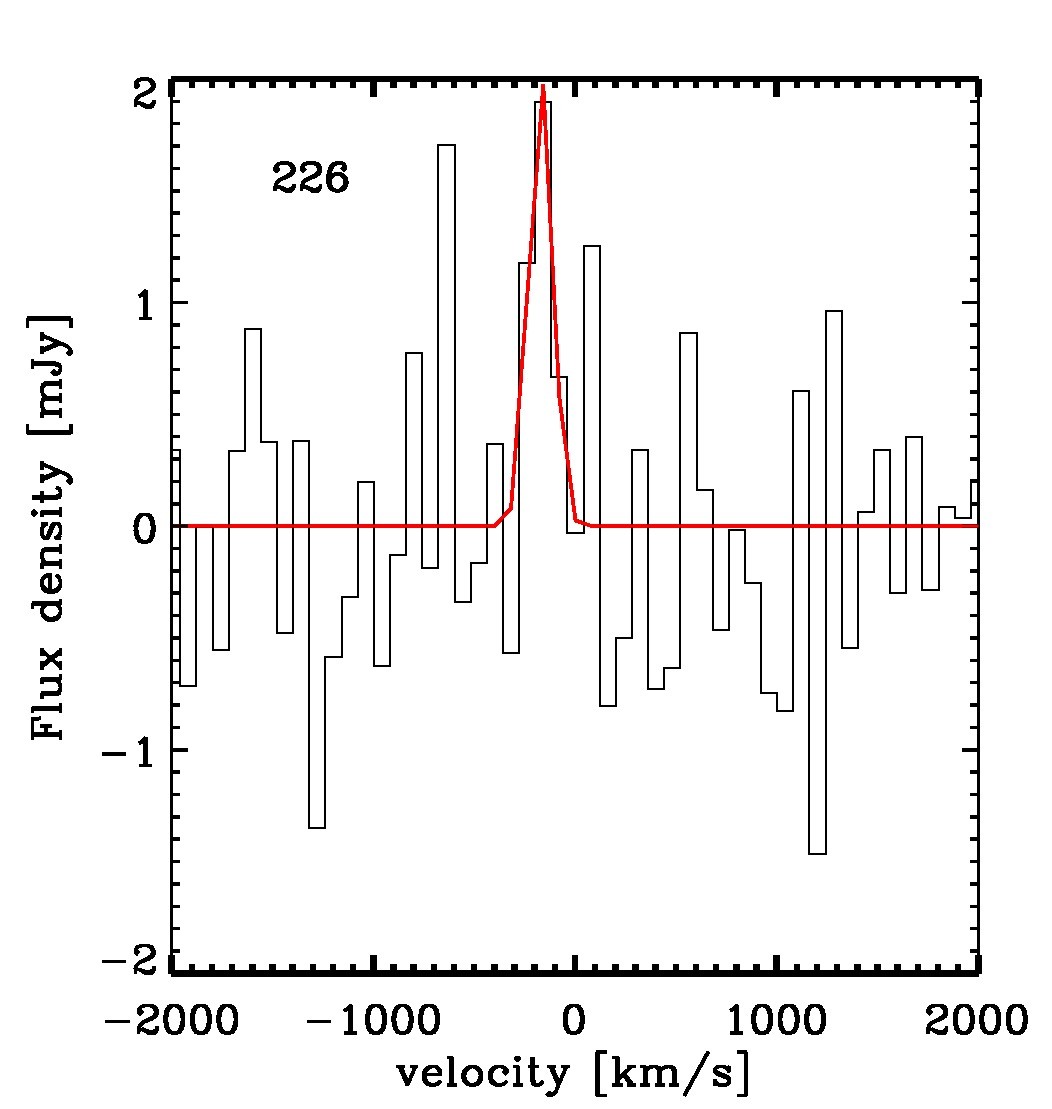}}
\caption{The Band 3 ALMA spectra of the three CO detected objects
  around the CO(2-1) line: C92 (left panel), C1591 (middle panel) and
  \#226 (right panel). The black line corresponds to the observed
  spectrum while the red line corresponds to the Gaussian fit. For
  C92, a single channel has a width of 50 km/s while for C1591 and
  \#226, the channel width is 80 km/s. The parameters corresponding to
  the Gaussian fits are given in Table \ref{table:fit_results}.
\label{fig:spectra}}
\end{figure*}

\begin{figure*}
\subfloat{\includegraphics[width=5in]{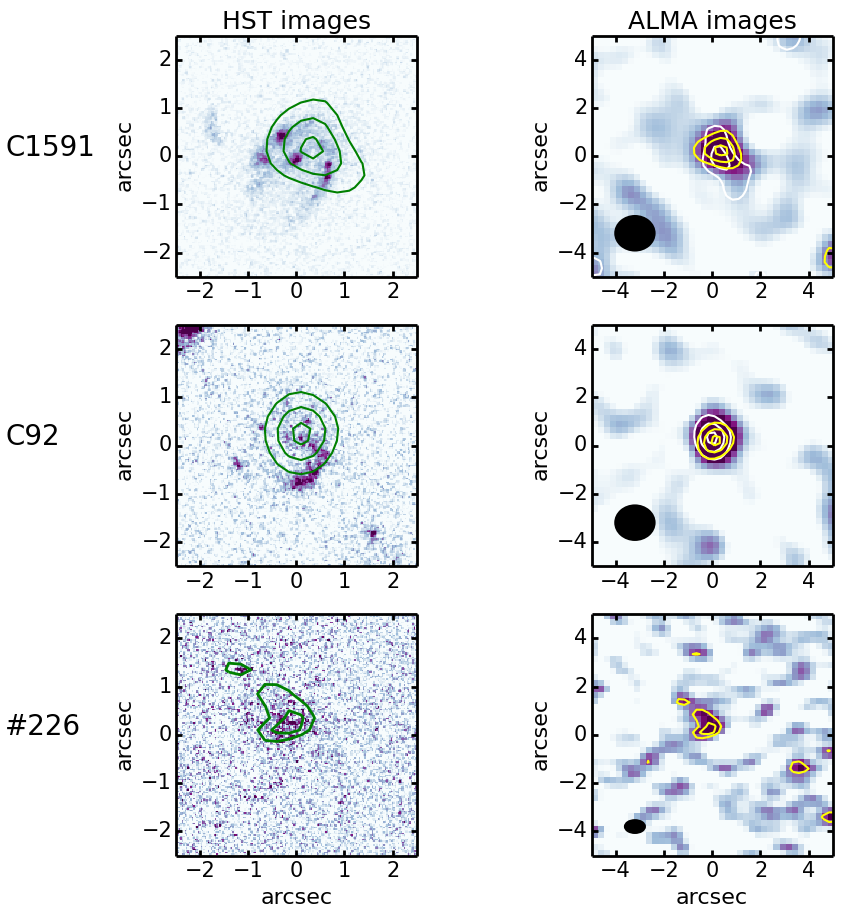}} 
\caption{{\it Left panels:} HST-ACS $5^{\prime\prime} \times 5^{\prime\prime}$ images of C1591 (top), C92 (middle) and \#226 (bottom). The overlaid CO contours are at levels 0.6, 0.9 and 1.1 Jy km/s over a velocity bin of 880 km/s for C1591, 0.5, 0.7 and 0.9 Jy km/s over a velocity bin of 350 km/s for C92 and 0.2, 0.3 and 0.4 Jy km/s over a velocity bin of 320 km/s for \#226. All the HST images show that the molecular gas traced by the CO emission line has a similar spatial extension compared to the optical/UV counterpart. {\it Right panels:} CO(2-1) emission line integrated maps over velocity bins of 880 km/s (C1591), 350 km/s (C92) and 320 km/s (\#226). The overlaid yellow and white contours in C1591 and C92 represent the CO emission in the two peaks visible in the respective spectrum in Fig. \ref{fig:spectra} while the yellow contours in \#226 represent the CO emission from the entire line. The CO emission in C1591 is marginally resolved. The inset ellipse is the synthesized beam during observations.
\newline Note the different scales of the two panels, for visualization purposes. North is up and East is towards left in all the maps.
  \label{maps}}
\end{figure*}

\section{Results and Analysis} \label{sect:results}

We detect CO(2-1) emission in 3 out of the 10 targets observed with
ALMA. C92 and C1591, both MS outliers, are detected with a >5$\sigma$ and $\sim$4$\sigma$ significance respectively, while \#226, in the CDFS field and a MS candidate, is
detected with $\sim$3$\sigma$ significance. The reported significances
are observed both in the spectra and the moment (flux) maps. The
statistics were obtained around a $25^{\prime\prime}$ region with
respect to the field center. The results from our analysis are
reported in Table \ref{table:observed_target_properties}.

\begin{table*}
\centering
\begin{tabular}{cccccccc}
\hline
ID & $\mathrm{z_{CO}}^{a}$ & $\mathrm{t_{exp}}^{b}$ & Beam size$^{c}$ & $\mathrm{I_{CO}}^{d}$ & Log $\mathrm{L^{\prime}_{CO}}^{e}$ & $\mathrm{M_{H_{2}}}^{f}$ & $\mathrm{f_{gas}}^{g}$\\
 & & (s) & $^{\prime\prime}\times ^{\prime\prime}$ & (Jy km/s) & (K km/s pc$^{2}$) & (M$_{\odot}$) &\\
\hline
DETECTIONS & & & & & & & \\
C1591*  & 1.237 & 151.2  & 1.6x1.4 & 0.86$\pm$0.21 & 10.2$\pm$2.4 & 10.7$\pm$2.6 & 0.2$\pm$0.1\\
C92*    & 1.581 & 151.2  & 1.6x1.4 & 1.01$\pm$0.13 & 10.4$\pm$1.2 & 11.0$\pm$1.3 & 0.4$\pm$0.1\\
\#226   & 1.389 & 151.2  & 0.8x0.5 & 0.29$\pm$0.11 & ~9.8$\pm$3.7 & 10.4$\pm$3.9 & 0.2$\pm$0.2\\
NON-DETECTIONS & & & & & & & \\
X5308 & - & 151.2  & 1.6x1.4 & <0.60 & <10.0 & <10.6 & <0.3\\
X2522 & - & 151.2  & 1.6x1.4 & <0.65 & <10.2 & <10.8 & <0.2\\
C1148 & - & 151.2  & 1.6x1.4 & <0.65 & <10.2 & <10.8 & <0.3\\
C488  & - & 483.84 & 1.5x1.4 & <0.30 & ~<9.6 & <10.2 & <0.5\\
C152  & - & 483.84 & 1.5x1.4 & <0.35 & ~<9.7 & <10.3 & <0.4\\ 
C103  & - & 483.84 & 1.5x1.4 & <0.30 & ~<9.8 & <10.4 & <0.3\\ 
\#682   & - & 151.2  & 0.8x0.5 & <0.60 & ~<9.9 & <10.5 & <0.5\\
\hline
\end{tabular}
\caption{Observed target properties:\newline
$^{a}$Redshift calculated from the observed CO(2-1) peak, \newline
$^{b}$On source exposure time in seconds, \newline
$^{c}$Beam size achieved during observations, \newline
$^{d}$The CO flux in Jy km/s. In case of non-detection this represents a 3$\sigma$ upper limit on the flux obtained from sensitivity reached over 240 km/s channel width around the expected position of CO line (see text for further details). \newline
$^{e}$The CO(1-0) luminosity calculated following \citet{solomon05}. We employ a conversion factor of 0.8 to convert from $\mathrm{L_{CO(2-1)} ~to ~L_{CO(1-0)}}$. The values of $\mathrm{L^{\prime}_{CO}}$ for non-detection represent the 3$\sigma$ upper limit. \newline
$^{f}$Molecular mass calculated from CO luminosity following the canonical relation, $\mathrm{M_{H2} = \alpha_{CO}L^{\prime}(CO)}$ with $\alpha_{CO}=3.6$, commonly used in literature for MS galaxies, \newline
$^{g}$Gas fraction $\mathrm{f_{gas}=M_{H_{2}}/(M_{H_{2}} + M_{\ast})}$. \newline
*All these galaxies are classified as starbursts according to the MS line from \citet{schreiber15} as mentioned in Table \ref{table:target_properties}.
  \label{table:observed_target_properties}}
\end{table*}

During the design of our ALMA program we used the parametrization of
the MS line for star forming galaxies presented in \citet{pannella09}
which is shown by the blue line in
Fig. \ref{fig:sample_selection}. All targets but C92 fell uniformly
along this line. For C92 we have in the meantime revised the estimate
of the SFR which moved it outside the MS region. As mentioned in
Sect. \ref{sect:target_selection}, we adopt here the latest
parametrization of the MS from \citet{schreiber15} as it traces the
higher mass end of the slope more accurately. Consequently, the slope
of the entire MS line is flatter than the one quoted by
\citet{pannella09}\footnote{Due to low number statistics at the 
high mass end, it is debated if the slope is constant throughout this 
range of mass.}. This resulted in C92 and C1591 being classified as
starburst galaxies.

The extracted spectra of each detected object with the Gaussian fits
to the emission line profiles are shown in Fig. \ref{fig:spectra} and
the results of various fitting parameters are reported in Table
\ref{table:fit_results}. In all cases, the peak of
the CO emission is blueshifted compared to the optical wavelengths
obtained from the COSMOS \& CDFS catalogue, which is apparent from the
central velocity of the Gaussian fits as reported in Table
\ref{table:fit_results}. Redshifts obtained from the optical
catalogues in the respective fields have been used as a reference for
all the plots.

\begin{table}
\centering
\begin{tabular}{cccc}
\hline
ID & v$_{center}^{a}$ & $\Delta$v$^{b}$ & f$^{c}$ \\
& (km/s) & (km/s) & (Jy km/s)\\
\hline
C1591 & -133.36$\pm$80 & 616$\pm$71 & 0.86$\pm$0.21\\
C92   & -77.23$\pm$50  & 224$\pm$18 & 1.01$\pm$0.13\\
\#226   & -169.80$\pm$80 & 137$\pm$22 & 0.29$\pm$0.11\\
\hline
\end{tabular}
\caption{Line fitting results for the three detections: $^{a}$Central velocity of the Gaussian fit, $^{b}$Width (full width half maximum [FWHM]) of the Gaussian, $^{c}$Flux of the Gaussian.
\label{table:fit_results}}
\end{table}

\subsection{Molecular mass from CO}
C1591 has a very broad CO emission line profile with a FWHM of $616\pm 71$
km/s as shown in Fig. \ref{fig:spectra}, middle panel. The CO emission
from C1591 shows a double peaked profile suggesting a rotating
galaxy. The flux map corresponding to the integrated CO profile is shown
in Fig. \ref{maps}, top right panel. The overlaid white and yellow
contours are the emission from these two peaks, each of width 200
km/s. The top left panel of Fig. \ref{maps} shows the HST-ACS image
of C1591 with CO emission line contours of the entire profile
overlaid suggesting a similar extension of CO emission compared to the UV/optical image. Using CASA task {\it imfit}, we fit an elliptical Gaussian model
to the CO(2-1) emission line integrated map over a velocity width of 880 km/s which gives values of the image component size convolved with the beam as 2.25$\pm$0.15$^{\prime\prime}$ and 1.51$\pm$0.16$^{\prime\prime}$. This suggests a marginal extension in the CO emission profile in the image. Deconvolving the image from the beam, we arrive at a half-light radius of $\sim$0.5$^{\prime\prime}$ ($\equiv$4.15 kpc) for C1591.

The CO emission line width in C92 is narrower with respect to C1591
and with a high significance of detection. Therefore, we were able to get the
spectrum in smaller bins 50 km/s wide (see
Fig. \ref{fig:spectra}). A closer look suggests a double peaked CO
emission profile in C92 as well and we constructed the flux maps with
the overlaid contours, similar to the case of C1591. This is shown in
the middle panel of Fig. \ref{maps}. The HST map does not reveal a wider extent of molecular gas in the host galaxy compared to the UV/optical image and the CO map shows that the emission is not resolved with the synthesized ALMA beam. An
X-shooter spectrum of C92, which is also known as XID60053, is
available and the results were published in \citet{brusa15}. This
object has a high extinction ($\mathrm{A_{V}\sim 6}$) due to which the
$\mathrm{H_{\alpha}}$ and [OIII] emission lines are barely detected.

The marginally resolved CO emission from C1591 can be used to
calculate the gas mass using the dynamical mass method which has been
frequently used in literature \citep[e.g.][]{carilli10, daddi10,
  hodge12, tan14}. Within a half-light radius ($\mathrm{r_{1/2}}$),
the dynamical mass for a spherically symmetric case is given by the
following equation:

\begin{equation}
M_{dyn}(r < r_{1/2})\simeq \frac{5\sigma^{2}r_{1/2}}{G}
\end{equation}
\noindent
where $\mathrm{\sigma = \Delta v_{CO}/2.355}$ is the velocity
dispersion of the CO emission line with $\mathrm{\Delta v_{CO}}$ being
the line width (FWHM) and $G$ is the gravitational constant. The gas
mass can then be derived by subtracting the stellar mass and the dark
matter component using the following equation \citep{daddi10}:

\begin{equation}
M_{dyn} = 0.5\times (M_{\ast} + M_{gas}) + M_{DM}(r < r_{1/2})
\end{equation}

Here we assume a dark matter component within the half-light radius to
be 25\% of the total dynamical mass \citep[e.g.][]{daddi10}. This
assumption is based on observations of local spirals \citep{pizagno05}
where the total mass at the half-light radius is dominated by
baryons. Using this dynamical method, we obtain a gas mass of
$\mathrm{6.0\pm2.4\times 10^{10} M_{\odot}}$ for C1591. At the redshift of
C1591, HI gas content is negligible compared to $\mathrm{H_{2}}$,
hence the reported gas mass is believed to be the mass of molecular
hydrogen in the host galaxy \citep[e.g.][]{neeleman16}.

For all sources, including C1591, we calculate the CO luminosity using
the result from \citet{solomon05}:

\begin{equation}
L^{\prime}_{CO} = 3.25 \times 10^{7} ~I_{CO} ~\nu_{obs}^{-2} ~D_{L}^{2} ~(1+z)^{-3}
\end{equation}
\noindent
where $\mathrm{I_{CO}}$ is the CO[2-1] flux in Jy km/s, $\mathrm{\nu_{obs}}$
is the observing frequency in GHz, $\mathrm{D_L}$ is the luminosity distance in Mpc at 
redshift $z$. In case of non-detection, we use the rms from the measurements (see
Sect. \ref{sect:observations_reductions} for a description on
calculation of rms) to estimate 3$\sigma$ upper limits on the CO
luminosity. Stacking of spectra from the sources with no detection revealed 
no emission lines, justifying our estimation of upper limits for these sources.
In all cases, we assume an excitation correction of 0.8 to
convert from $\mathrm{L_{CO(2-1)}}$ to
$\mathrm{L_{CO(1-0)}}$\footnote{Literature values range from
  0.7-0.85. See
  e.g. \citet{frayer08,daddi10,brusa15,silverman15}}. The calculated
CO luminosities and the upper limits are reported in Table
\ref{table:observed_target_properties}.

The above calculated CO luminosities are then converted into molecular
mass using $\mathrm{M_{H_{2}} =
  \alpha_{CO}L^{\prime}(CO)}$.
$\mathrm{\alpha_{CO}}$ presents the biggest systematic uncertainty in
molecular gas measurements using CO since its value could lie anywhere
between 0.8-3.6 \citep[e.g.][]{arimoto96,bolatto13} depending on
various factors such as metallicity and specific SFR (sSFR) \citep[e.g.][]{genzel12, hunt15}. For the
purpose of our study, we use an $\mathrm{\alpha_{CO}}$ value of 3.6,
which is commonly used in literature for MS galaxies at a similar
redshift as our AGN sample
\citep[e.g.][]{daddi10, tacconi13}, ignoring the possibility that AGNs might have different $\mathrm{\alpha_{CO}}$ due to hard ionizing radiation (see Sect. \ref{sect:discussion}). The gas mass
measurement using the dynamical method for C1591 described above gives
an $\alpha_{CO}$ value of 4.1$\pm$1.6 which is well within the value of 3.6 
used here for all cases. To be consistent, we recalculate the molecular gas mass for  the MS comparison sample using $\mathrm{\alpha_{CO}}$=3.6 and for literature data using CO[3-2] emission line, we employ an excitation correction of 2 to convert to CO[1-0] luminosity following \citet{tacconi13}. 

\subsection{Observed relations for AGNs}

The main goal of this paper is to identify whether the presence of an
AGN affects the molecular gas content of its host galaxy and
consequently the star formation process. To answer these questions, we
compare our molecular mass measurements with those of normal star
forming galaxies and starburst galaxies without an AGN. Details of
the comparison sample were mentioned earlier in
Sect. \ref{sect:target_selection}. We stress again that
  the comparison sample has been matched in redshift and stellar mass
  to make sure we compare galaxies with similar properties.

\begin{figure*}
\subfloat{\includegraphics[width=6.0in, height=5.0in]{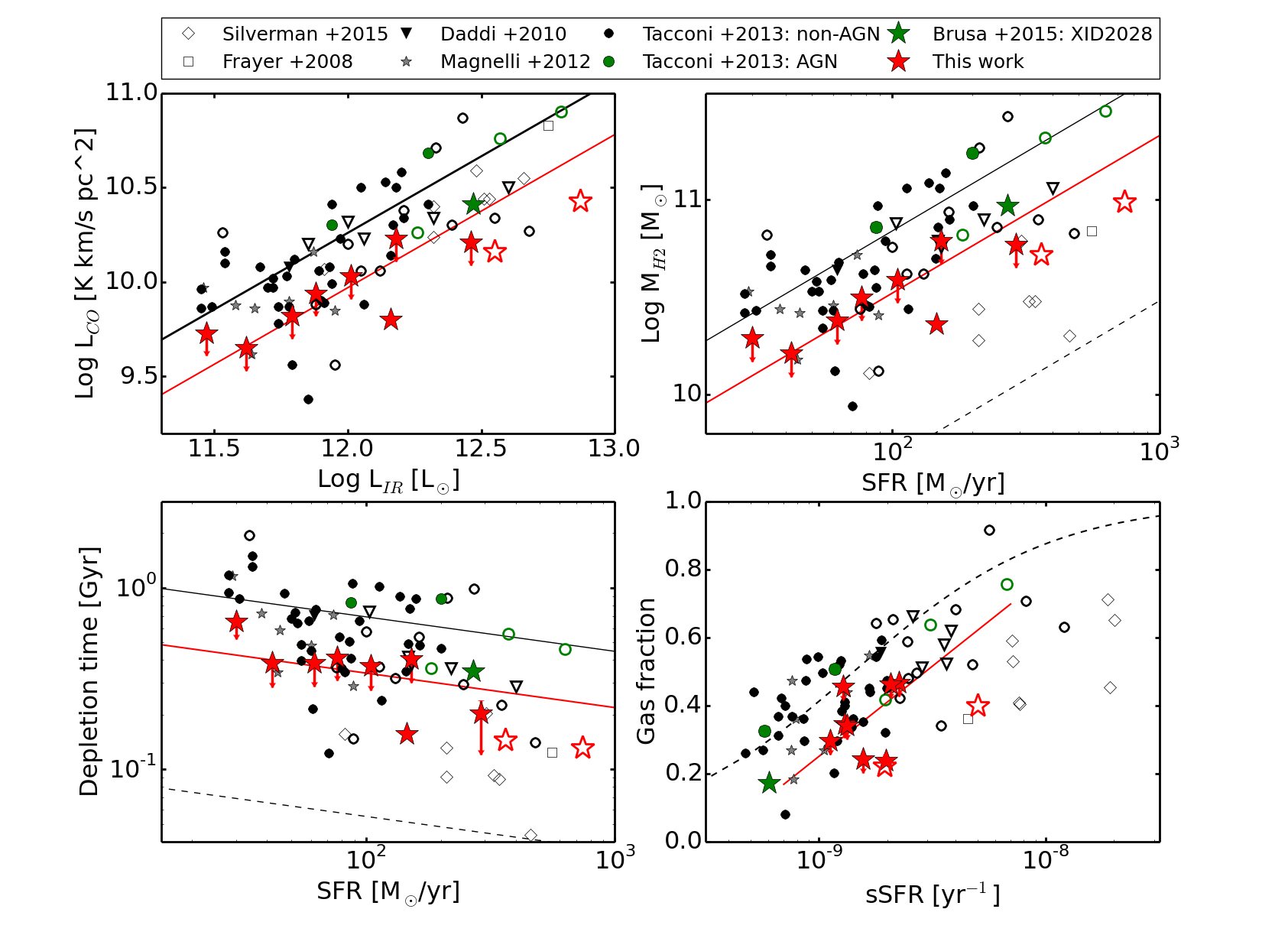}} 
\caption{In all plots, the colored symbols represent the AGN sample from this work (red stars-filled and unfilled) and the previous works (green symbols). The filled symbols represent the main sequence galaxies while the open symbols represent the starburst population. The red line represents the best fit for AGNs in the MS, considering all the 3$\sigma$ upper limits as detections. {\it Top left panel:} CO luminosity vs. total infrared luminosity. The solid black line is the correlation adapted from \citet{daddi10}. {\it Top right panel:} CO based molecular gas mass measurements vs. SFR for the entire sample. We assume a conversion factor of $\alpha_{CO}$ to be 3.6. In our ALMA sample, we have one detection from the MS galaxies and two detections from the SB galaxies. The black solid line and the dashed line are the correlations between $\mathrm{M_{mol}}$ and SFR for MS galaxies and SB galaxies respectively, adapted from \citet{sargent14}. {\it Lower left panel:} The depletion time (=$\mathrm{M_{H_{2}}/SFR}$) vs. SFR plot for the entire sample. The solid and the dashed lines are the positions of the MS and SB galaxies as compiled in \citet{sargent14}. {\it Lower right panel:} Gas fractions vs. specific SFR (=SFR/M$_{\ast}$) for the entire sample. The dotted line shows the position of main sequence non-AGN galaxies adapted from \citet{tacconi13}.
  \label{fig:analysis_plots}}
\end{figure*}

Figure \ref{fig:analysis_plots} shows the various correlations
obtained for our targets in the context of the entire comparison
sample. Before fitting any function to the AGN sample, we
did a survival two-sample test using R function {\it survdiff()}
implemented in python on each plot to check the probability that the
AGN sample and the non-AGN sample are drawn from the same
distribution. The function returns a $p$-value from a chi-squared
distribution. For all the correlations derived in this
section, we obtain a $p$-value of 0.008 i.e. a >99\% probability that
our AGN sample has a different distribution from the normal MS
star forming galaxies. While deriving the correlations for
AGNs, we kept the slope of the linear function to be equal to that of
the MS relations for ease of comparison with the non-AGN sample.

The top left panel in Fig. \ref{fig:analysis_plots} shows
  $\mathrm{L^{\prime}_{CO}}$ vs. $\mathrm{L_{IR}}$ plot for galaxies
  between redshift 1 and 2. The infrared luminosity, $\mathrm{L_{IR}}$
  here represents the total infrared luminosity for all the sample in
  the plot. The color coding of different
points is the same as in Fig. \ref{fig:sample_selection}. We employ a
standard excitation correction of 0.8 to convert from CO(2-1)
luminosity to CO(1-0) luminosity for our as well as all the comparison
samples for consistency. First of all, we do not find a significant
difference in the $\mathrm{L^{\prime}_{CO}}$-$\mathrm{L_{IR}}$
correlation for MS and non-AGN starburst galaxies as reported in
previous works such as those of \citet{genzel10}, \citet{sargent14}, and \citet{tacconi13}. These works used a compilation of local as well
as high redshift galaxies. However, when we only consider the high
redshift galaxies between 1$<z<$2, only one correlation is required to
describe the star burst and MS population in the
$\mathrm{L^{\prime}_{CO}}$-$\mathrm{L_{IR}}$ plane given by the
following mean equation adapted from \citet{sargent14} for MS
galaxies and shown as a black line in Fig. \ref{fig:analysis_plots},
upper left panel:

\begin{equation}
Log ~L^{\prime}_{CO} [L{\odot}] = (0.81\pm0.03)\cdot Log ~L_{IR}[L{\odot}] + (0.54\pm0.02)
\end{equation}
\noindent
This is consistent with the more recent results \citep[e.g.][]{genzel15, silverman15}.
Our MS AGN sample lie below this correlation and considering all the
3$\sigma$ upper limits as detections, we derive an average best-fit function
for $\mathrm{L^{\prime}_{CO}}$-$\mathrm{L_{IR}}$ correlation for AGNs:
\begin{equation}
Log ~L^{\prime}_{CO}[L{\odot}] = 0.81\cdot Log ~L_{IR}[L{\odot}] + (0.25\pm0.35)
\end{equation}
\noindent
This is given by the red line in Fig. \ref{fig:analysis_plots}, upper
left panel. Thus the mean $\mathrm{L^{\prime}_{CO}}$-$\mathrm{L_{IR}}$
correlation for AGNs is a factor ``at least'' 2 lower than that of
main sequence non-AGN galaxies. Such a trend is also recovered in a flux-flux plot as well. We also stress here that the derived line for AGNs is itself an upper limit since most of the data points used to fit this correlation are upper limits.

The difference in the properties of the two sets of galaxies (non-AGN star forming and the starburst galaxies) becomes
apparent once we convert these CO luminosities into gas masses and
this is shown in Fig. \ref{fig:analysis_plots}, top right panel. The
solid and the dotted lines are the average correlations obtained by
\citet{sargent14}, where the dotted line indicates the 
limiting locus for starbursts 
which are strong outliers to the MS and have among the largest star formation 
efficiencies (SFEs). This plot is an ideal 
indicator of the SFE. The galaxies 
that follow a lower
correlation, such as the starburst galaxies along the dotted black line, 
have a higher SFE compared to those following a higher correlation,
such as the MS galaxies along the solid black line. This is because
for the same molecular gas mass, there is more star formation in the
SB galaxies compared to MS galaxies. Our AGN host galaxies on the MS
are consistently below the average MS line in the
$\mathrm{M_{H_{2}}-SFR}$ plane indicating that the star formation
efficiency of the AGN host galaxies is higher than for the 
non-AGN population. We derive the functional form for the best fit
relation for the AGNs in the MS (taking the upper limits as
detections):

\begin{equation}
Log ~M_{H_{2}} [M_{\odot}] = 0.81\cdot Log ~SFR [M_{\odot}/yr] + (8.90\pm0.35)
\end{equation}

The intercept for this equation is a factor of $\sim$2 lower
than the average line for the MS galaxies (see \citet{sargent14} for the
best fit $\mathrm{M_{gas}-SFR}$ relation for main sequence 
non-AGN galaxies).
 These values thus support
our assertion above that AGN host galaxies tend to have a higher SFE
compared to M$_{\ast}$ and SFR matched non-AGN galaxies. We
note here that the $\alpha_{CO}$ value adopted for our galaxies is the
same as for the main sequence non-AGN population used in this study. In a scenario where hard ionizing radiation
  from AGNs could destroy these molecules, conversion factor
  $\alpha_{CO}$ would be lower for AGN host galaxies and the estimated
  gas mass would decrease even further. This argument is further strengthened once we take into
account the single detection we have from ALMA which is well below the
locus of the MS galaxies. In fact, this detection falls in the
region of the starburst galaxies in the $\mathrm{M_{H_{2}}-SFR}$ 
plane at this redshift. The AGN host
galaxies which are starbursts also fall into the same region which
confirms previous results that starbursting AGN hosts have properties
similar to those of the non-AGN starburst population.

Lower gas mass for a given star formation rate is indicative of higher
SFE as discussed above and lower depletion time scales as well. This
is confirmed by the depletion time scale ($t_{dep}=\mathrm{M_{gas}/SFR}$)
vs. SFR plot in the lower left panel of
Fig. \ref{fig:analysis_plots}. Starbursts tend to occupy the region
with lower gas depletion time scales as they are
believed to be efficient in converting the available cold gas content
into stars, while in the case of normal star forming galaxies, they occupy
a region of higher $\mathrm{t_{dep}}$ (the exact functional forms of the
two lines are given in \citet{sargent14}). The best fit relation
between $\mathrm{t_{dep}}$ and SFR for our AGN sample in the MS is:

\begin{equation}
Log ~t_{dep} [Gyr] = -0.19*Log ~SFR [M_{\odot}/yr] - (0.09\pm 0.35)
\end{equation}
\noindent
Our AGN sample on the MS have depletion time scales at least a factor of 2 lower 
than those of the main sequence non-AGN galaxies of similar stellar mass.

Finally, for each one of our AGNs we derived the gas fraction,
$\mathrm{f_{gas}=M_{gas}/(M_{gas} + M_{\ast})}$ and plotted it against
the specific SFR (sSFR) as shown in the lower right panel of
Fig. \ref{fig:analysis_plots}. The gas fraction rather than the total
gas content is the most appropriate parameter to use in order to
compare the gas content of galaxies independently of their stellar
masses. The best fit correlation for AGN host galaxies can be
described using the following equation:

\begin{equation}
f_{gas} = 0.53\cdot Log ~sSFR + (5.02\pm 0.35)
\end{equation}

This is shown by the solid red line in Fig. \ref{fig:analysis_plots},
lower right panel and the relation is valid for
$\mathrm{10^{-9}~yr^{-1} < sSFR< 10^{-8}~yr^{-1}}$. The black dashed line in the
same panel shows the best fit function for MS galaxies adapted from
\citet{tacconi13}. Although some of our data points fall within the
general population of the MS galaxies, the average line fitting
function for AGNs show gas fractions less by a factor of at least 2
compared to the MS galaxies. Hence, together with a higher SFE, these
moderate luminosity AGN host galaxies also show a lower gas fraction.

\section{Discussion}\label{sect:discussion}

The observation of high SFE, lower gas depletion time scales and lower
gas fractions in our AGN sample may be an indirect evidence of AGN
feedback on the host galaxy. However with the current data we cannot 
distinguish between a positive or a negative feedback scenario (in 
other words, we cannot tell if the AGNs are shifted 
right or down in Fig. \ref{fig:analysis_plots}).

{\it Negative Feedback:} As mentioned earlier, AGNs are known to host
powerful outflows within their host galaxies. These outflows may drive
the gas away from the galaxy, hence preventing further star formation
without affecting the current star formation rate. As an example,
for XID2028, it has been shown from observations with SINFONI
\citep{cresci15} that this galaxy hosts a powerful outflow traced by
the ionized gas. The PdBI molecular gas observation for this object
using CO(3-2) emission line is shown by the green star in all the
plots \citep{brusa15}. Similar to the properties of our AGN sample,
this target seems to have a higher SFE, lower depletion time scale and
lower gas fraction compared to the ``average'' best fit functions
of the MS galaxies. A confirmation of the depleted gas content in
XID2028 comes from the molecular gas mass measurements obtained from the dust
mass, which is a factor of $\sim$2 lower than the values obtained assuming
$\mathrm{\alpha_{CO}=3.6}$, supporting the 
hypothesis that the outflow observed in the ionized gas
phase may be depleting the cold molecular gas in the host galaxy.

To prove the presence of on-going outflows in our sample, 
we will need deeper observations in the sub-mm and near
infrared regime to trace the molecular and ionized gas 
phase respectively. The presence of
molecular gas observations coupled with the outflow power estimates in
such a sample will tell us whether there is a
relation between the outflow power and the available molecular gas in
the host galaxy.

{\it Positive feedback:} An alternative scenario is that the AGN could in fact enhance
the star formation as well which would result in an increase in the SFE of the host
galaxies. Incidences of such positive AGN feedback have been reported
in literature \citep[e.g.][]{croft06,silk13,gabor14,cresci15a}
and it is believed that the star formation is triggered in the
pressurized medium as the outflows shock against the surrounding
ISM. The effects of positive feedback could be investigated
with high resolution 0.1" continuum and CO imaging, coupled with
similar resolution star formation maps.

{\it Change in gas phase due to AGN}: If the outflows are not powerful enough to expel the gas, the strong ionizing radiation from AGNs might also have the ability to change the status of the gas by either destroying the molecules or ionizing them. This could be observed in other wavebands of the electromagnetic spectrum. Thus, a multi-wavelength approach is fundamental to understand the observations of the AGN sample presented in this paper.

{\it Starburst followed by an AGN phase}: Lastly, the current observations would also be consistent with a starburst episode followed by an AGN phase \citep{bergvall16} and we might happen to catch the targets in the latter phase. An insight into the star formation history of the AGN sample might give us a clue about the observed correlations.

Finally it is worth noting that the AGN sample from \citet{tacconi13} does not
show any remarkable difference in its molecular gas properties from
those of the non-AGN population at odds 
with our sample and the target from
\citet{brusa15a}. For example, in Fig. \ref{fig:analysis_plots},
lower right panel, the gas fraction of the \citet{tacconi13} AGN
sample fall almost on (or a bit above) the MS line as while our
sample falls below this MS line by a factor $> 2$ on average.
The main difference between our AGN and the AGN sample from
\citet{tacconi13} is that our AGN are on average brighter
($\mathrm{L_{bol} \sim 10^{43}-10^{47}}$ erg/s) than the PHIBBS AGN
(i.e. the \citet{tacconi13} AGN) which have X-ray luminosity of
$\mathrm{10^{42}-10^{43.5}}$ erg/s which translates into
$\mathrm{L_{bol} \sim 10^{43.5}-10^{45}}$ erg/s. We infer that in the case
of our AGNs, the gas might have been depleted by a more powerful AGN feedback
due to the higher bolometric luminosities, which are responsible for the
observed lower gas fraction.

\section{Summary and Conclusions} \label{sect:summary}

We have presented the molecular gas mass measurements obtained from ALMA observations of $\mathrm{^{12}CO[2-1]}$ in a representative sample of 10 AGNs, 8 of which lie on the MS of star forming galaxies and 2 are
classified as starbursts. Their selection is based on their position
 with the aim to compare the molecular gas fractions and SFE
of these AGN hosts on the MS with a redshift, M$_{\ast}$ and SFR
matched sample of non-AGN star forming galaxies. The following
points summarize results and conclusions of this work:

\begin{itemize}
\item We detect CO emission in 3 out of the 10 AGNs observed with
  ALMA, C92, C1591 and \#226 with a significance of
    >5$\sigma$, $\sim$4$\sigma$ and $\sim$3$\sigma$ respectively. The
  emission line widths are in the range $\sim$140-620 km/s. For the
  rest of the sample, we use 3$\sigma$ upper limits as proxy for CO
  luminosity for analysis.
  
\item Molecular gas for all the objects has been
  determined assuming the conventional conversion factor,
  $\alpha_{CO}$ to be 3.6. For C1591, a starbursting AGN system, the
  observed CO emission is marginally resolved and we could use
  dynamical analysis as well to measure the gas mass. The gas mass
  measured using the dynamical analysis agrees within the
  errors with that estimated using
  $\mathrm{\alpha_{CO}=3.6}$.

\item Using the same conversion factor between CO luminosity and
  molecular gas mass for the non-AGN comparison
  sample and the AGN sample,we find that AGN hosts have
    on average a gas fraction lower by a factor of $\ga$2, higher SFR
    and consequently $\ga$2.3 times lower gas depletion time scales
    compared to non-AGN main sequence galaxies of similar stellar
    mass and SFR at z$\sim$1.5. In fact, the AGNs follow a different
  correlation compared to the non-AGN comparison 
  sample. These differences
  are significant at the $> 99\%$ level. 

\item Our hypothesis is that the gas depletion may be an indirect evidence of the impact of AGN feedback on the host galaxy. Powerful winds from the central source might be
    sweeping the galaxies clean of molecular gas content and/or
    enhance star formation in the shocked medium. The AGN radiative
    field might also change the status of the ISM by heating,
    ionization and disruption of molecules. Multiwavelength studies of
    statistical samples, aimed to prove or disprove this scenario are
    ongoing.
\end{itemize}

\section*{Acknowledgements}
We thank G. Popping and M. Bethermin for helpful discussions and
suggestions. M. Brusa acknowledges support from the FP7 Career
Integration Grant "eEASY: supermassive black holes through cosmic time
from current surveys to eROSITA-Euclid Synergies" (CIG 321913). SC
acknowledges financial support from the Science and Technology
Facilities Council (STFC). CF acknowledges funding from
  the European Union's Horizon 2020 research and innovation programme
  under teh Marie Sklodowska-Curie Grant agreement No 664931.  MTS acknowledges support from a Royal Society Leverhulme Trust Senior Research Fellowship (LT150041). This paper makes use of the following ALMA
data: ADS/JAO.ALMA\#2013.1.00171.S. ALMA is a partnership of ESO
(representing its member states), NSF (USA) and NINS (Japan), together
with NRC (Canada), NSC and ASIAA (Taiwan), and KASI (Republic of
Korea), in cooperation with the Republic of Chile. The Joint ALMA
Observatory is operated by ESO, AUI/NRAO and NAOJ.

%%%%%%%%%%%%%%%%%%%%%%%%%%%%%%%%%%%%%%%%%%%%%%%%%%



\bibliographystyle{mnras}
\bibliography{reference.bib} % if your bibtex file is called example.bib

%%%%%%%%%%%%%%%%%%%%%%%%%%%%%%%%%%%%%%%%%%%%%%%%%%

%%%%%%%%%%%%%%%%% APPENDICES %%%%%%%%%%%%%%%%%%%%%

%\appendix



%%%%%%%%%%%%%%%%%%%%%%%%%%%%%%%%%%%%%%%%%%%%%%%%%%

% Don't change these lines
\bsp	% typesetting comment
\label{lastpage}
\end{document}